# Dynamics of a Service Economy Driven by Random Transactions

Robert W. Easton
Professor Emeritus, Department of Applied Mathematics
University of Colorado
Robert.Easton@colorado.edu

**Abstract:** Agents buy and sell services. All services are of equal quality. Buyers choose sellers at random. Monetary and fiscal policies are imposed by a central bank and a central government. Credit is supplied by a commercial banking system. Propensities to buy, sell, and lend depend on account balances, interest rates, tax rates and loan default rates. Computer simulations track weekly sales, loans, account balances, commercial bank profits, solvency and compliance with reserve requirements, and government debt.

The model of this economy is fully specified by a computer program. The program allows the user to explore the effects of parameter changes. Monetary and fiscal policies are implemented by choices of parameters such as interest rates and reserve requirements, and government tax and spending rates. Credit supply, consumer confidence, and loan default rates strongly affect the behavior of the economy in terms of sales.

## *Random Transactions*

Agents and transactions form the foundation for this model of a service economy. A potential transaction occurs between a buyer, a seller, and a lender. The odds of a transaction (or sale of service) occurring are affected by opportunity, willingness to buy, willingness to sell, money, credit, and trust (among buyer, seller, lender). Sellers are assumed to offer service of equal quality. To generate transactions buyers choose sellers at random. The sale occurs with a probability that depends on the buyer's willingness to buy, the seller's willingness to sell, and the bank's willingness to lend.

Instead of using cash, the agents make electronic transfers of credits and debits between their accounts. In a typical transaction agent *j* buys service from agent *k*. The bank is notified and it debits account *j* and credits account *k*. The government collects a tax on each transaction. The buyer pays for *x* hours of service, the seller gets $(1 - tax)xp$ fiat dollars, and the government gets $(tax)xp$ fiat dollars, where *p* is the price of one hour of service.

The price of service is fixed at $p = 1$ fiat dollar per one hour of service for the basic model discussed here. Market prices can vary and several economies with separate currencies can trade with each other in more complicated models.

One can create models of service economies driven by random transactions as computer programs and then run the programs to see how the models work. The effects of policy choices such as changes in tax rates, interest rates, and government spending can be explored.

The actors in the model discussed here consists of a government, a central bank, a representative commercial bank, and agents exchanging services. The central bank keeps two accounts: one for the commercial bank and one for the government. The government account has no debt limit. The central bank sets the interest rate on government debt. The government buys services and collects taxes from agents.

The commercial bank is required by the central bank to keep a fraction $R=\rho D$ of its deposits $D$ as reserves in its account at the central bank. The central bank does not pay interest on reserves. The funds available to the commercial bank for loans are the owner's capital $K$ plus deposits minus reserves. Thus outstanding loans $L$ must satisfy the inequality $L \leq K + D - \rho D$. A further solvency requirement is also imposed. The bank's capital $K$ must be positive.

The commercial bank keeps accounts for the agents. It issues loans and collects interest. (In principle, competition with other banks or government regulation would keep the commercial bank from charging excessive interest rates.) It buys service from agents and it may also buy government debt. The commercial bank transfers taxes on transactions to the government account.

The commercial bank loan policy in this model depends on account balances and the risk of defaults on loans. The bank stops lending when the total amount loaned is constrained by the reserve requirement. This prevents all agents with negative accounts from buying on credit until bank reserves increase.

The commercial bank pays interest on deposits (positive accounts) equal to the interest rate on government debt set by the central bank. Negative accounts pay the loan interest to the commercial bank. Both the commercial bank and the government buy services. The commercial bank invests all deposits that are not held as reserves in T-bills so that effectively the government pays the interest on these deposits and the bank collects the interest on loans.

Fiscal Policy: The government sets tax and spending rates. For example the tax rate might be 20% and the spending rate 22% of the gross weekly product. In this case the government runs a fiscal deficit.

Monetary Policy: The central bank sets the interest rate on government debt and the reserve requirement. Directly through the reserve requirement and indirectly, by buying and selling government debt, the central bank controls the credit supply. It acts to promote full employment (or gross weekly sales).

Risk: To introduce risk in the model, choose a default probability $D(A)$ that decreases as the balance in an account increases. To be specific, the model sets a default probability of one (certainty of default) for accounts below a default limit. The default probability decreases linearly to a default rate of zero for positive accounts. Each week the agent with the lowest account balance is at risk of defaulting. If the agent's account is negative, roll dice using the default probability to decide if a default occurs. If it does, the account is reset to zero and a new agent takes over. The amount of the default is subtracted from the commercial bank account. The bank limits risk by refusing to loan to accounts below a loan limit. The default limit and the loan limit are parameters in the model.

## Model Simulations

The following is a list of parameters used by the model. The comment after the percent sign explains the parameter. The computer code is included in appendix A. Each line of code is explained so that a reader does not need to know the programming language to understand the algorithm.

```
N=10; %number of agents
W=53; %number of weeks
T=10; %number of weekly transactions
rl = .07/52; % interest rate on loans charged by the bank
spend= 0.0;%banker's spending policy as a fraction of loan interest
rd= 0.06/52; % interest rate on deposits paid by the government
tax= 0.20; %government's tax rate on transactions
spendtaxes=1.0; %government spends a multiple of taxes collected
mood= 7;% odds of buying on credit out of 10
defaultlimit= -500;%Balances below this limit always default
loanlimit = -5; %bank won't loan to accounts below this limit
```

Experiment 1. Average weekly sales on three simulation runs were 6.7925, 6.4151, and 6.5094 out of a possible 10. The sales and loans for each week from the first run are shown in the bar graphs below. The third figure shows a chart of the commercial banks' account at the central bank. The fourth figure shows a chart of compliance with the reserve requirement. A default occurred around week 30 and shows up as a loss in the bank's account and also as a dip in the compliance graph. While the bank was not insolvent, it was out of compliance (the graph of the compliance variable must be positive) and could not make loans for the next week.

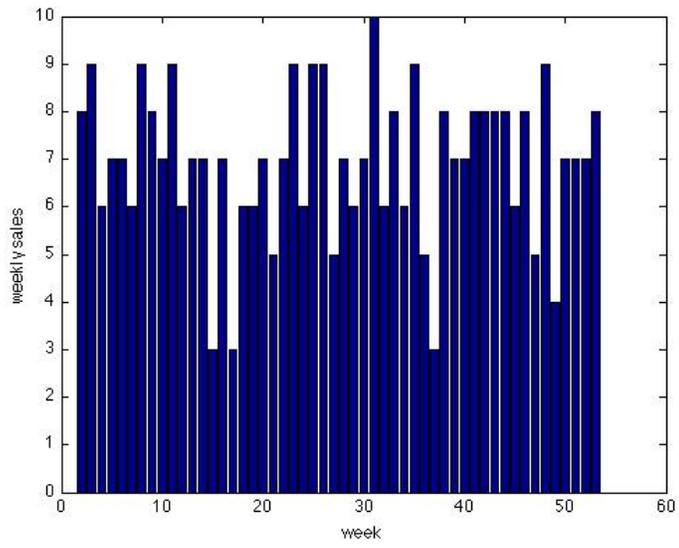

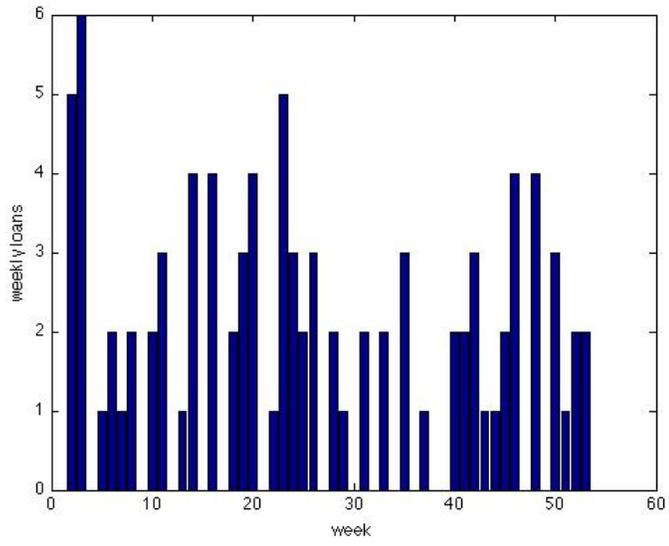

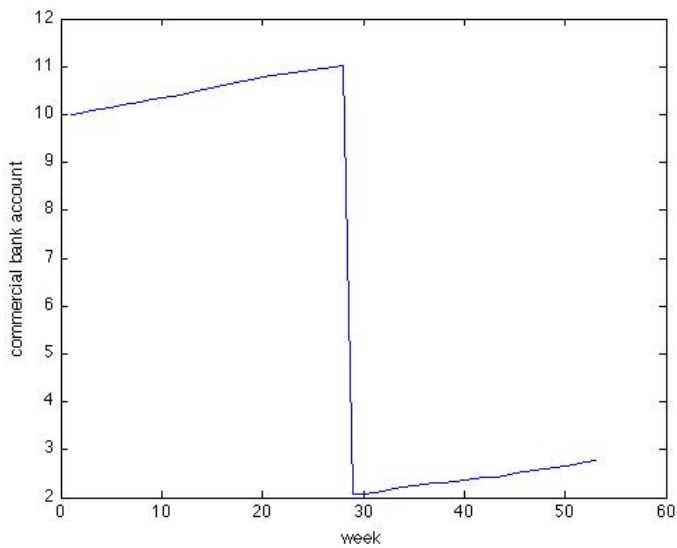

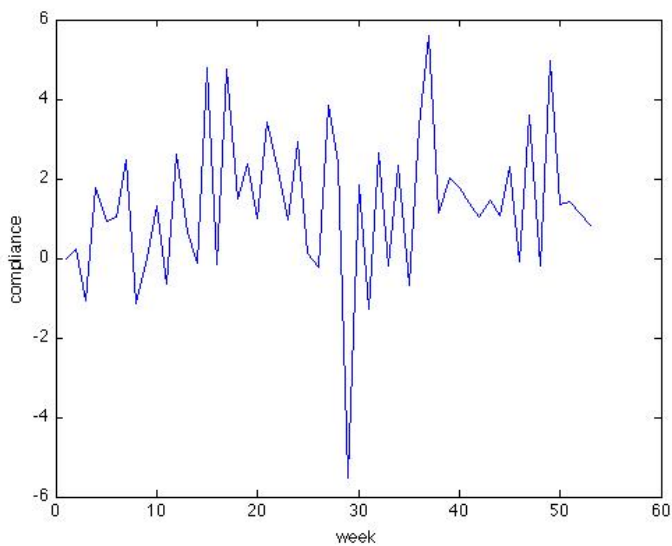

Experiment 2. Setting the loan limit at –15, the bank took on more risk. Weekly sales averaged 6.5660, 6.7925, and 6.8868. On the first run defaults occurred at weeks 30 and 44. The bank was insolvent after the second default and stopped making loans.

It is hard to keep the bank solvent in this model. The outcomes of simulation runs are quite sensitive to the rates of default, and the loan limit set by the bank. By limiting credit the bank limits losses. Of course it also makes smaller loans and less profit.

## Retrospective

Note that the tax rate does not directly affect the employment rate in this model because the government spends taxes to hire service equally from all agents. Deficit spending by the government increases employment and sales. A high tax rate could produce more negative accounts and thus fewer buyers. It could also decrease the desire to sell. So far we assume the seller always agrees to sell. Financial inequality could have negative effects if the increase in consumption by the rich was more than canceled by the decrease in consumption by the poor.

The government is able to spend more than it collects in taxes. An interesting experiment using the model is to find the level of government spending that is sustainable at a fixed tax rate and borrowing costs. This depends also on the initial government debt.

Cutting taxes does not create more jobs in this model, rather it shifts purchase decisions to the private sector. Income from service sold by agents is diminished by the taxes they pay. The government spends the tax revenue to provide public services. This points to the crux of the political division over taxes. The cost and benefit of a private purchase is evident to the buyer. The cost of government is evident to the taxpayer while the benefits are shared by the entire society. The rich may believe that they pay more in taxes than they receive in government services.

In the model the sellers pay taxes. This is like an income tax. An alternative is to tax the buyers. This would be like having a consumption tax. One could also split the tax between buyers and sellers.

## Market Prices

So far the price and quantity of service have been constant. If the price of service is allowed to vary, buyers will want to buy more service as the price declines. Sellers will want to provide more service as the price increases. In theory, the price adjusts to equalize buyers demand for service with the supply offered by sellers.

For simplicity, assume that all agents have the same propensities to buy and to sell. Assume that a buyer's and a seller's price versus quantity functions are simple linear functions. The variable $h$ is the quantity of service in hours and the variable $A$ is the balance in an agent's account. As a buyer, the agent's price versus quantity equation is
$$p^b = 2 - K[h - eA].$$
As a seller, the agent's price versus quantity equation is
$$p^s = K[h + eA].$$

When both buyer and seller have zero balances, the choice of parameters $K=1/5$ and $e=1/10$ produces the price of 1 fiat dollar for one hour of service, and the quantity 5 hours of service as the common solution to both the buyer's and seller's price versus quantity equations. The parameter $e$ determines the sensitivity of the buyer and seller to changes in account balances.

Equating the buyer's demand and the seller's supply of service determines a quantity-price combination that is satisfactory to both. The balances in the buyer's and in the seller's accounts may differ. Thus if agent $i$ is the buyer and agent $j$ is the seller, then we have supply equal to demand when $2 - K[h - eA_i] = K[h + eA_j]$, where $A_i$ is the balance in the buyer's account and $A_j$ is the balance in the seller's account. The quantity-price amounts agreeable to both parties are $h_{i,j} = 1/K + e[A_i - A_j]/2$, $p_{i,j} = 1 + Ke[A_i + A_j]/2$.

One could use these amounts to conclude a transaction. Instead the computer program of a model with market prices computes the average price over all "transaction opportunities" during a week, and uses the average price as a market price. Thus the market price *p* is the average of the prices $p_{i,j}$ taken over all buy-sell pairs picked during the week. Now all transactions occur at the market price with varying quantities of service. Agent *i* receives $h_i = 2/K + eA_i - p/K$ hours of service and pays the amount $ph_i$ to agent *j*.

## Conclusions

The use of random transactions combined with computer programs provides a new foundation for creating macro-economic models. The model presented here is an example of this process. It is possible to use this foundation to create models of international trade between multiple economies, each with their own currencies, governments, and central banks. Further research is directed at building models that incorporate capital, and multiple goods and services.

## Appendix A: Computer Code (Using Matlab software)

```
%% Financial Model
% Components: a commercial bank, a central bank, a government, agents buying and
%selling services. Buyers choose random sellers.
%%Parameters
N=10; %number of couples in the co-op
W=53; %number of weeks
T=10; %number of weekly transactions
rl =.07/52; % interest rate on loans charged by the Bank
spend=.0;%banker's spending policy as a fraction of loan interest
rd=0.06/52; % interest rate on deposits paid by Gov
tax=.20; %government's tax rate on transactions
spendtaxes=1.0; %government spends a multiple of taxes collected
mood=7;% odds of buying on credit out of 10
defaultlimit=-500;%Balances below this limit always default
loanlimit = -5; %bank won't loan to accounts below this limit
%% Initialize Variables
A=0*ones(1,N); % initial bank accounts for each agent
All=0*ones(W,N); %a matrix to store the weekly accounts
WeeklySales=0*ones(1,W); %number of sales for each week
CB=0*ones(1,W); %commercial bank account at the central bank
CB(1)=10; %initial reserves of the commercial bank
G=0*ones(1,W);% the government account at the central bank
sales=0;%count weekly sales
```

```matlab
loans=0;%count number of loans
compliance=0*ones(1,W); %required ratio of reserves to deposits
%% Random Transactions
for week=2:W
   sales=0; %count sales for one week
   loans=0; %count loans for one week
   for i=1:N %Each buyer chooses a seller at random
   seller(i)=unidrnd(N); %choose a seller at random
   % If buyer and seller are different end, else repeat
   u=i-seller(i);
   while u==0
   seller(i)=unidrnd(N);
   u=i-seller(i);
   end
   end %end of buyer-seller selection

%% create N potential transactions
   for i=1:N %decide odds to buy
   if A(i)>=loanlimit && A(i)<=0 &&compliance(week-1)>=0 && CB(week-1)>=0
   %if the bank has required reserves and is solvent and the borrower is
   %not too far in debt
      coin=unidrnd(10); %choose a random integer from 1-10
      if coin<=mood %request a loan
      % the commercial bank always loans when compliance>=0
         A(i)=A(i)-5; %debit buyer for 5 hours of service
         A(seller(i))=A(seller(i))+(1-tax)*5;%credit seller
         sales=sales+1;%count a sale
         loans=loans+1; %count a loan
      end
    end
   if A(i)<10 && A(i)> 0 % if the account is between 0 and 10
      coin=unidrnd(10); %set odds of a buy
      if coin<=9 %first couple buys 9 out of 10 times
       A(i)=A(i)-5; %debit buyer for 5 hours of service
         A(seller(i))=A(seller(i))+(1-tax)*5;%credit seller
         sales=sales+1;%count a sale
      end
   end
   if 10<A(i) %first couple always buys
        A(i)=A(i)-5; %debit buyer for 5 hours of service
        A(seller(i))=A(seller(i))+(1-tax)*5;%credit seller
        sales=sales+1;%count a sale
   end %end of the transaction
   end %end of the N transactions for the current week
WeeklySales(week)=sales;%WeeklySales(week)=total sales for the week
Weeklyloans(week)=loans;
```

```matlab
%% Update the commercial bank account
Neg=.5*(A-abs(A)); % this sets positive accounts to zero
L=-Neg*ones(N,1); %L is the total amount of loans (for current week)
Pos=.5*(A+abs(A)); % this sets negative accounts to zero
D=Pos*ones(N,1); %total deposits in agents accounts
%at this point A=Neg+Pos
LoanInterest= -Neg*rl; %interest on loans (vector)
DepInterest=Pos*rd; % interest on deposits (vector)
A=A-LoanInterest+DepInterest; %member account balances updated
CB(week)=CB(week-1)+rl*L;% credit commercial bank account with loan interest
A=A+spend*rl*L*ones(1,N)/N; %the bank buys service equally from all agents
CB(week)=CB(week)-spend*rl*L; %update the commercial bank account

%% Update the Government account
taxrevenue=tax*sales*5; % credit government account with weekly tax revenue
G(week)=G(week-1)+taxrevenue; %add tax revenue for the current week
G(week)=G(week)-rd*D; %deduct deposit interest from government account
% the government buys service equally from all agents
A=A+spendtaxes*taxrevenue*ones(1,N)/N;
G(week)=G(week)-spendtaxes*taxrevenue; %debit government account
All(week,:)=A; %store the weekly account balances in the matrix All

%% Default Risk
%choose the lowest account and default with
%a probability y that depends of the debt. Set the account back to zero.
[B,I]=min(A); %choose the minimum account and save its index I
% for the defaultlimit see the parameter list
if B<defaultlimit %default is certain if B < defaultlimit
    y=1;
elseif B>0 % positive accounts never default
    y=0;
elseif defaultlimit <=B % defaults decline as B increases
    y=(B/defaultlimit);
end
%choose a number x in [0,1] from the uniform distribution. If x<y default
%set the balance in account I to zero. Use rand(1) command to pick x.
if rand(1)<y;% if a default occurs
    All(week,I)=0;%set minimum account back to zero
    A(I)=0; %erase debt from current week account
    CB(week)=CB(week)+B; %deduct loss from CB account
end
compliance(week)=CB(week)+.9*D-L; %update the reserve requirement
end %end of for week =2:W  loop

%% Display Graphs and Print Results
```

```
All; % display the matrix containing agent's accounts
AveWeeklySales=(1/W)*WeeklySales*ones(1,W)'% Average weekly sales
AveBankAccount=CB*ones(W,1)/W; %Average weekly commercial bank account
figure(1)
bar(WeeklySales) %plot weekly sales as a bar graph
xlabel('week'); ylabel('weekly sales')
figure(2)
plot(G)%plot government account
xlabel('week'); ylabel('government account')
figure(3)
plot(CB)
xlabel('week'); ylabel('commercial bank account')
figure(4)
plot(compliance)
xlabel('week'); ylabel('compliance')
figure(5)
bar(Weeklyloans) %plot weekly loans
xlabel('week'); ylabel('weekly loans')
```

## Acknowledgements:

Thanks to Professor Martin Boileau for teaching me macro-economics, for listening to my ideas, and for providing valuable feedback. Thanks also to Dr. John Flint, and Dr. Stuart Ambler for valuable conversations and comments. This work was originally inspired by Paul Krugman's discussions of a babysitting co-op (see *Baby-sitting the Economy*, Slate Magazine, 8/13/98). The references listed below helped shape my thinking on modeling financial systems.